\documentclass[9pt]{article}
\usepackage{spconf}
\usepackage{amsmath,amssymb}
\usepackage{graphicx}
\usepackage{verbatim}
\usepackage{multirow}
\usepackage{cite}

\title{Prime Factor Cyclotomic Fourier Transforms 
  with Reduced Complexity over Finite Fields}

\name{Xuebin Wu, Ning Chen, and Zhiyuan Yan}
\name{Xuebin Wu, Zhiyuan Yan, Ning Chen, and Meghanad Wagh}
  
\address{Department of ECE,  Lehigh University,
  Bethlehem, PA 18015\\
  E-mails: \{xuw207, yan, nic6, mdw0\}@lehigh.edu}

\renewcommand{\L}{\mathbf{L}}
\newcommand{\f}{\mathbf{f}}
\newcommand{\Q}{\mathbf{Q}}
\newcommand{\C}{\mathcal{C}}
\newcommand{\A}{\mathbf{A}}

\begin{document}
\maketitle
\begin{abstract}
  Discrete Fourier transforms~(DFTs) over finite fields have
  widespread applications in error correction coding.  Hence, reducing
  the computational complexities of DFTs is of great significance,
  especially for long DFTs as increasingly longer error control codes
  are chosen for digital communication and storage systems. Since DFTs
  involve both multiplications and additions over finite fields and
  multiplications are much more complex than additions, recently
  proposed cyclotomic fast Fourier transforms (CFFTs) are promising
  due to their low multiplicative complexity. Unfortunately, they have
  very high additive complexity. Techniques such as common
  subexpression elimination (CSE) can be used to reduce the additive
  complexity of CFFTs, but their effectiveness for long DFTs is
  limited by their complexity. In this paper, we propose prime factor
  cyclotomic Fourier transforms (PFCFTs), which use CFFTs as sub-DFTs
  via the prime factor algorithm. When the length of DFTs is prime,
  our PFCFTs reduce to CFFTs. When the length has co-prime factors,
  since the sub-DFTs have much shorter lengths, this allows us to use
  CSE to significantly reduce their additive complexity. In comparison
  to previously proposed fast Fourier transforms, our PFCFTs achieve
  reduced overall complexity when the length of DFTs is at least
  $255$, and the improvement significantly increases as the length
  grows. This approach also enables us to propose efficient DFTs with
  very long length (e.g., 4095-point), first efficient DFTs of such
  lengths in the literature. Finally, our PFCFTs are also advantageous
  for hardware implementation due to their regular structure.
\end{abstract}

\section{Introduction}
Discrete Fourier transforms~(DFTs) over finite fields have widespread
applications in error correction coding, which in turn is used in all
digital communication and storage systems. For instance, both syndrome
computation and Chien search in the syndrome based decoder of
Reed-Solomon (RS) codes, a family of error control codes with
widespread applications, can be formulated as polynomial evaluations
and hence can be implemented efficiently via DFTs over finite
fields. Implementing an $N$-point DFT directly requires $O(N^2)$
multiplications and additions, and becomes costly when $N$ is
large. Hence, reducing the computational complexities of DFTs has
always been of great significance. Recently, efficient long DFTs
become particularly important as increasingly longer error control
codes are chosen for digital communication and storage systems. For
example, RS codes over GF$(2^{12})$ and with block length of several
thousands are considered for hard drive \cite{IDEMAWhitePaper} and
tape storage \cite{Han2005} as well as optical communication systems
\cite{Buerner2004} to achieve better error performance; syndrome based
decoder of such RS codes requires DFTs of length up to $4095$ over
GF$(2^{12})$. Furthermore, regular structure of DFTs is desirable as
it is conducive to efficient hardware implementation.

For DFTs over the complex field, many techniques have been proposed to
reduce the computational complexity, leading to various fast Fourier
transform (FFTs). Prime factor algorithm (PFA) \cite{Good1960} and
Cooley-Turkey algorithm (CTA) \cite{Cooley1965} can implement an
$N$-point DFT with $O(N\log{N})$ multiplications for $N$ with a lot of
small factors. The PFA was applied to DFTs over finite fields
\cite{Truong2006}, but DFTs obtained via the PFA still have high
multiplicative complexity. In contrast, recently proposed cyclotomic
FFTs (CFFTs) \cite{Trifonov2003} are promising due to their low
multiplicative complexity. Based on efficient algorithms for short
cyclic convolutions, CFFTs require much fewer multiplications at the
expense of very high additive complexity.  Properly designed common
subexpression elimination (CSE) algorithms (see, for example,
\cite{Chen2009c}) can greatly reduce the additive complexity of CFFTs
for short and moderate lengths, but they are much less effective for
long DFTs. This is because the run time and storage requirement of the
CSE algorithm in \cite{Chen2009c} become infeasible for large lengths
(say $2047$ or $4095$). As a result, a simplified and less effective
CSE algorithm was used to reduce the additive complexity of 2047-point
CFFTs in \cite{Ning2009}, but the additive complexity of the
2047-point CFFTs in \cite{Ning2009} remains very high.  This
complexity issue results in a lack of \textbf{efficient} DFTs of very
long lengths in the literature: to the best of our knowledge, the
CFFTs in \cite{Ning2009} is the only 2047-point DFTs, and efficient
4095-point DFTs cannot be found in the literature.  An additional
disadvantage of CFFTs is their lack of structure and regularity, which
makes it difficult to implement CFFTs in hardware efficiently.

In this paper, we propose prime factor cyclotomic Fourier transforms
(PFCFTs), which use CFFTs as sub-DFTs via the prime factor
algorithm. When the length of DFTs is prime, our PFCFTs reduce to
CFFTs. When the length has co-prime factors, since the sub-DFTs have
much smaller lengths, this allows us to use CSE to significantly
reduce their additive complexity. In this case, although out PFCFTs
have slightly higher multiplicative complexity than CFFTs, they have
much lower additive complexity. As a result, our PFCFTs achieve
smaller overall complexity than \textbf{all} previously proposed FFTs
when the length of DFTs is at least $255$, and the improvement
significantly increases as the length grows. This approach also
enables us to propose efficient DFTs with very long length (e.g.,
4095-point), first efficient DFTs of such lengths in the
literature. Our PFCFTs also have a regular structure, which is
suitable for efficient hardware implementations. Although the PFA is
also used in \cite{Truong2006}, our work is different in two ways: (1)
the sub-DFTs are implemented by CFFTs; (2) CSE is used to reduce the
additive complexity of DFTs. The reduced complexity of our PFCFTs is a
result of these two differences.

The rest of the paper is organized as
follows. Section~\ref{sec:background} briefly reviews the necessary background to make this paper self-contained. 
In Section~\ref{sec:PFCFT} we propose our PFCFTs, and compare their complexity with previously proposed FFTs. The advantage of our PFCFTs  in hardware implementation is discussed in Section~\ref{sec:arch}. Concluding remarks are provided
in Section~\ref{sec:conclusion}.

\section{Backgrounds}
\label{sec:background}

\subsection{Cyclotomic fast Fourier transforms}
Let $\alpha \in $ GF$(2^l)$ be a primitive $N$-th root of 1 (this implies that $N | 2^l-1$, otherwise $\alpha$ does not exist). Given an
$N$-dimensional vector $\mathbf{f}=(f_0,f_1, \cdots, f_{N-1})^T$ over
GF$(2^l)$, the DFT of $\mathbf{f}$ is given by $\mathbf{F}=(F_0, F_1,
\cdots, F_{N-1})^T$, where $F_k=\sum_{n=0}^{N-1}f_n\alpha^{nk}$.

It is shown in \cite{Trifonov2003} that the DFT is given by $\mathbf{F}=\mathbf{A}\mathbf{L}\mathbf{\Pi}\mathbf{f}$,
where $\mathbf{A}$ is an $N\times N$ binary matrix, $\mathbf{\Pi}$ is
a permutation matrix, $\mathbf{L}=\mathrm{diag}(\L_0, \L_1, \cdots,
\L_{m-1})$ is a block diagonal matrix with square matrices $\L_i$'s 
on its diagonal, and $m$ is the number of cyclotomic cosets modulo $N$
with respect to GF$(2)$. The $i$-th block $\L_i$ is an $m_i\times m_i$
circulant matrix corresponding to a cyclotomic coset of size $m_i$,
which is generated from a normal basis $\{\gamma_i^{2^0},
\gamma_i^{2^1}, \cdots, \gamma_i^{2^{m_i-1}}\}$ of GF$(2^{m_i})$, and
is given by
\begin{equation*}
  \L_i=\left[
    \begin{array}{cccc}
      \gamma_i^{2^0}      & \gamma_i^{2^1} & \cdots & \gamma_i^{2^{m_i-1}}\\
      \gamma_i^{2^1}      & \gamma_i^{2^2} & \cdots & \gamma_i^{2^{0}}\\
      \vdots             & \vdots        & \ddots & \vdots         \\
      \gamma_i^{2^{m_i-1}} & \gamma_i^{2^0} & \cdots & \gamma_i^{2^{m_i-2}}
    \end{array}
  \right].
\end{equation*}
Let $\f'=\mathbf{\Pi}\f=(\f_0'^T, \f_1'^T, \cdots, \f_{m-1}'^T)^T$,
and $\f_i'$ has a length of $m_i$.  The multiplication
between $\L_i$ and $\f_i'$ can be formulated as an $m_i$-point cyclic
convolution between $\mathbf{b}_i=(\gamma_i^{2^0},
\gamma_i^{2^{m_i-1}}, \gamma_i^{2^{m_i-2}}, 
\cdots, \gamma_i^{2^{1}})^T$ and $\f_i'$. Since $m_i$ is usually
small, using efficient bilinear algorithms
for short cyclic convolutions, $\L_i\f_i'$ can be computed efficiently by
$$\L_i\f_i'=\mathbf{b}_i \otimes\f_i'=\mathbf{Q}_i(\mathbf{R}_i\mathbf{b}_i\cdot\mathbf{P}_i\f_i')
= \mathbf{Q_i}(\mathbf{c}_i\cdot\mathbf{P}_i\f_i'),$$ where
$\mathbf{P}_i$, $\mathbf{Q}_i$, and $\mathbf{R}_i$ are all binary
matrices, $\mathbf{c}_i=\mathbf{R}_i\mathbf{b}_i$ is a precomputed
constant vector, and $\cdot$ denotes an entry-wise multiplication
between two vectors. Combining all the matrices, we get CFFTs
\begin{equation}
  \mathbf{F}=\mathbf{A}\mathbf{Q}(\mathbf{c}\cdot\mathbf{P}\f'),
  \label{eq:cfft}
\end{equation}
where both $\Q=\mathrm{diag}(\Q_0,\Q_1, \cdots, \Q_{m-1})$ and $\mathbf{P}=\mathrm{diag}(\mathbf{P}_0,
\mathbf{P}_1,\cdots, \mathbf{P}_{m-1})$ are block
diagonal matrices. 

The only multiplications needed in (\ref{eq:cfft}) are entry-wise multiplication $\mathbf{c}\cdot\mathbf{P}\f'$, and the multiplications of binary matrices $\mathbf{A}$, $\mathbf{Q}$, and $\mathbf{P}$ with vectors
require only additions. Implemented directly, CFFTs in (\ref{eq:cfft}) require much fewer multiplications than direct implementation, at the expense of very high additive complexity.

\subsection{Common subexpression elimination}
Common subexpression elimination is often used to reduce the additive complexity of a collection of additions.
Consider a matrix-vector multiplication between an $N\times M$ binary
matrix $\mathbf{M}$ and an $M$-dimensional vector $\mathbf{x}$ over a
field $\mathbb{F}$. It can be done with additive operations only, the
number of which is denoted by $\mathcal{C}(\mathbf{M})$ since the
complexity is determined by $\mathbf{M}$ and irrelevant with
$\mathbf{x}$. It has been shown that minimizing the number of additive
operations, denoted by $\mathcal{C}_\mathrm{opt}(\mathbf{M})$, is an
NP-complete problem \cite{Carey1979}. Therefore it is almost
impossible to design an algorithm with polynomial complexity to find
the minimum number of additions.

Instead of finding an optimal solution, different algorithms have been
proposed to reduce $\mathcal{C}(\mathbf{M})$. The CSE algorithm
proposed in \cite{Chen2009c} takes advantage of the \emph{differential
  savings} and \emph{recursive savings}, and greatly reduces the
number of additions in calculating $\mathbf{M}\mathbf{x}$, although
the reduced additive complexity, denoted by
$\mathcal{C}_\mathrm{CSE}(\mathbf{M})$, is not always the
minimum. Furthermore, the CSE algorithm in \cite{Chen2009c} is
randomized, and the reduction results of different runs are not the
same. Therefore in practice, we can run the CSE algorithm many times
and choose the best results.  Using the CSE algorithm in
\cite{Chen2009c}, the additive complexity and overall complexity of
CFFTs with length up to $1023$ are greatly reduced. It is more
difficult to apply the CSE algorithm in \cite{Chen2009c} to CFFTs of
longer length. This is because though the CSE algorithm in
\cite{Chen2009c} is an algorithm with polynomial complexity (it is
shown that it has an $O(N^4+N^3M^3)$ complexity), its runtime and
storage requirement become prohibitive when $M$ and $N$ are very
large, which occurs for long DFTs.

\subsection{Prime factor and 
  Cooley-Turkey algorithms}
The basic idea of both the PFA and the CTA is to first decompose an
$N$-point DFT into shorter sub-DFTs, and then construct the $N$-point
DFTs based on the sub-DFTs. The PFA assumes that $N$ contains at least
two co-prime factors, that is, $N=N_1N_2$, where $N_1$ and $N_2$ are
co-prime. For any integer $n \in \{0, 1, \cdots, N-1\}$, there is a
unique integer pair $(n_1, n_2)$ such that $0\le n_1 < N_1$, $0\le n_2
< N_2$, and $n=n_1N_2+n_2N_1\pmod{N}$.
For any integer $k \in \{0, 1, \cdots, N-1\}$, suppose $k_1 = k
\pmod{N_1}$ and $k_2=k\pmod{N_2}$,
where $0\le k_1 < N_1$ and $0 \le k_2 < N_2$. By the Chinese
Remainder Theorem (CRT), $(k_1, k_2)$ uniquely determines $k$, and $k$
can be represented by $k=k_1N_2^{-1}N_2+k_2N_1^{-1}N_1 \pmod{N}$,
where $N_1^{-1}N_1=1\pmod{N_2}$ and $N_2^{-1}N_2=1\pmod{N_1}$.

Let $\alpha$ be a primitive $N$-th root of 1. Substituting the
representation of $n$ and $k$ in $\alpha^{nk}$, we get
$\alpha^{nk}=(\alpha^{N_2})^{n_1k_1}(\alpha^{N_1})^{n_2k_2}$,
where $\alpha^{N_2}$ and $\alpha^{N_1}$ are primitive  $N_1$-th root and $N_2$-th
root of 1, respectively. The $k$-th element of the DFT
is given by
\begin{equation}F_k = 
  \underbrace{\sum_{n_1=0}^{N_1-1}\Big(\overbrace{\sum_{n_2=0}^{N_2-1}f_{n_1N_2+n_2N_1}\alpha^{N_1n_2k_2}}^{N_2-\mathrm{point\,
        DFT}}\Big)\alpha^{N_2n_1k_1}}_{N_1-\mathrm{point\, DFT}}.
  \label{eq:pfa}
\end{equation}
Hence, the $N$-point DFT is expressed based on $N_1$-point and
$N_2$-point sub-DFTs. By first carrying out $N_1$ $N_2$-point DFT and
then $N_2$ $N_1$-point DFT, the $N$-point DFT is derived. Note that
the $N_1$- and $N_2$-point DFTs can be further decomposed by the PFA,
if $N_1$ and $N_2$ have co-prime factors.

The CTA differs from the PFA in that the CTA does not assume the
factors of $N$ are co-prime. The CTA also use different index
representations of $n$ and $k$. Let $N=N_1N_2$, then $n=n_1+n_2N_1$,
where $0\le n_1 < N_1$ and $0\le n_2 < N_2$, and $k=k_1N_2+k_2$, where
$0\le k_1< N_1$ and $0\le k_2 < N_2$. The $k$-th element of the DFT is
given by
\begin{equation}
  F_k=\underbrace{\sum_{n_1=0}^{N_1-1}\Big(\overbrace{\sum_{n_2=0}^{N_2-1}f_{n_1+n_2N_1}\alpha^{N_1n_2k_2}}^{N_2-\mathrm{point\,DFT}}\Big)\alpha^{n_1k_2}\alpha^{N_2n_1k_1}}_{N_1-\mathrm{point\,DFT}}.
  \label{eq:cta}
\end{equation}
Compared with (\ref{eq:pfa}), (\ref{eq:cta}) has an extra term
$\alpha^{n_1k_2}$, which is called the twiddle factor and requires
extra multiplications. However, the advantage of the CTA is that it
can be used for arbitrary composite length, including prime powers to
which the PFA cannot be applied. The CTA is often very effective if
$N$ has a lot of small prime factors. For example, $N$-point DFTs by
the CTA require $O(N\log{N})$ multiplications if $N$ is a power of
2. However, for DFTs over finite field GF$(2^l)$, the DFT lengths are
either $2^l-1$ or its factors, and they do not have many prime
factors. In addition, the multiplicative complexity due to the twiddle
factor is not negligible for DFTs over finite fields. Hence, we focus
on the PFA in this paper.

\section{Prime Factor Cyclotomic Fourier Transforms}\label{sec:PFCFT}
\subsection{Difficulty with long CFFTs}
Consider an $N$-point DFT. Suppose there are $m$ cyclotomic cosets
modulo $N$ with respect to GF$(2)$, and the $i$-th coset consists of
$m_i$ elements. Suppose an $m_i$-point cyclic convolution requires
$\mathcal{M}(m_i)$ multiplications, then the total number of the
multiplicative operations of implementing the $N$-point DFT is given
by $\sum_{i=1}^m\mathcal{M}(m_i)$ and the number of the additive
operations is $\mathcal{C}(\mathbf{AQ})+\mathcal{C}(\mathbf{P})$. The
multiplicative complexity can be further reduced since some elements
in the vector $\mathbf{c}$ in (\ref{eq:cfft}) may be equal to~1.  We
may then apply CSE to the matrices $\mathbf{AQ}$ and $\mathbf{P}$ to
reduce $\mathcal{C}(\mathbf{AQ})$ and $\mathcal{C}(\mathbf{P})$,
respectively. Since $\mathbf{P}=\mathrm{diag}(\mathbf{P}_1,
\mathbf{P}_2, \cdots, \mathbf{P}_m)$ is a block diagonal matrix, it is
easy to see that
$\C_\mathrm{opt}(\mathbf{P})=\sum_{i=1}^m\C_\mathrm{opt}(\mathbf{P}_i)$. Thus
one can reduce the additive complexity of each $\mathbf{P}_i$ to get a
better result of $\C(\mathbf{P})$. The size of $\mathbf{P}_i$ is much
smaller than that of $\mathbf{P}$, and it is possible to run the CSE
algorithm many times to achiever a smaller additive
complexity. However, the matrix $\mathbf{AQ}$ does not have this
property, and the CSE algorithm has to be applied directly on this
matrix. When the size of $\A\Q$ is large, the CSE algorithm in
\cite{Chen2009c} requires a lot of time and memory so that it becomes
impractical.  In \cite{Ning2009}, the reduced complexity of 2047-point
DFT over GF$(2^{11})$ is given after simplifying the CSE algorithm at
the expense of performance loss. For the same reason, it is difficult
to reduce the complexity of 4095-point DFT over GF$(2^{12})$ by the
CSE algorithm in \cite{Chen2009c}.

\subsection{Prime factor cyclotomic Fourier transforms}
Instead of simplifying the CSE algorithm or designing other low
complexity optimization algorithms, we propose prime factor cyclotomic
Fourier transforms by first decomposing a long DFT into shorter
sub-DFTs and then implementing the sub-DFTs by CFFTs. We denote the
additive (or multiplicative) complexity of an $N$-point DFT over
GF$(2^l)$ as $\mathcal{K}(N)$, and the algorithm is denoted in the
subscription of $\mathcal{K}$. If $N$ can be decomposed as a product
of $s$ co-prime factors $N_1, N_2, \cdots, N_s$, we can use the PFA to
decompose the $N$-point DFT into $N_1$-, $N_2$-, $\cdots$, and
$N_s$-point DFTs. Suppose we use CFFTs to compute these sub-DFTs, the
additive (or multiplicative) complexity of the $N$-point DFT is given
by
\begin{equation*}
  \label{eq:PFAcomp}
  \mathcal{K}_\mathrm{PFCFT}(N)=\mathcal{K}_\mathrm{PFCFT}(\prod_{i=1}^sN_i)=\sum_{i=1}^s\frac{N}{N_i}\mathcal{K}_\mathrm{CFFT}(N_i).
\end{equation*}
If the additive (or multiplicative) of an $N$-point CFFT is $O(N^2)$,
the additive (or multiplicative) of the corresponding PFCFT is
$O(N\sum_{i=1}^sN_i)$, which can be further reduced by CSE.  Since CSE
is more effective for shorter CFFTs, the decomposition makes it easier
to reduce the additive complexity of long DFTs. In \cite{Truong2006},
the PFA is used to reduce the DFTs complexity, but the idea of CFFT is
not used.


\begin{table}[!htb]
  \centering
  \begin{tabular}{c|cccc}
    \hline
    $L$ & mult. & $\mathcal{C}_\mathrm{CSE}(\Q^{(L)})$ &
    $\mathcal{C}_\mathrm{CSE}(\mathbf{P}^{(L)})$ & total\\
    \hline
    2  & 1  & 2   & 1  & 3\\
    3  & 3  & 5   & 4  & 9\\
    4  & 5  & 9   & 5  & 14\\
    5  & 9  & 16  & 10 & 26\\
    6  & 10 & 21  & 11 & 32\\
    7  & 12 & 25  & 22 & 47\\
    8  & 19 & 35  & 16 & 51\\
    9  & 18 & 40  & 31 & 71\\
    10 & 28 & 52  & 31 & 83\\
    11 & 42 & 76  & 44 & 120\\
    12 & 32 & 53  & 34 & 87\\
    \hline
  \end{tabular}
  \caption{Complexities of short convolutions}
  \label{tab:convCmplx}
\end{table}

\begin{table}[!htb]
  \centering
  \begin{tabular}{c|ccc}
    \hline
    lengths & mult. & add. (scheme 1)  & add. (scheme 2)\\
    \hline
    3       & 1    & 6    & 6 \\
    5       & 5    & 21   & \textbf{17} \\
    7       & 6    & 31   & \textbf{24} \\
    9       & 11   & 51   & \textbf{48} \\
    11      & 28   & 106  & \textbf{86} \\
    13      & 32   & 125  & \textbf{100} \\
    15      & 16   & 90   & \textbf{80} \\
    17      & 38   & \textbf{153}  & 183  \\
    23      & 84   & \textbf{335}  & 407 \\
    31      & 55   & \textbf{338}  & 358 \\
    33      & 85   & \textbf{420}  & 438 \\
    35      & 75   & 407  & \textbf{304} \\
    45      & 90   & 481  & \textbf{415} \\
    51      & 115  & \textbf{641}  & 755 \\
    63      & 97   & \textbf{791}  & 1038 \\
    65      & 165  & 1093 & \textbf{883}  \\
    73      & 144  & \textbf{1498} & 1567 \\
    85      & 195  & \textbf{1602} & 1817 \\
    89      & 336  & \textbf{2085} & 4326\\
    91      & 230  & 1668 & \textbf{1418} \\
    93      & 223  & 1910 & \textbf{1408} \\
    117     & 299  & 2328 & \textbf{2015} \\
    \hline
  \end{tabular}
  \caption{The reduced complexity for CFFTs whose lengths are less
    than 200 and are factors of $2^l-1$, $4\le l \le 12$. }
  \label{tab:shortFFTs}
\end{table}

\begin{table}[!tbh]
  \centering
  \begin{tabular}{c|cccc}
    \hline
    Length & Decomposition & mult. & add. & total\\
    \hline
    15 & $3\times 5$ & 20 & 81 & \textbf{221} \\
    \hline
    \multirow{1}{*}{63}  & $9 \times 7$ & 131 & 552 & \textbf{1993} \\
    \hline
    \multirow{4}{*}{255}
    & $3 \times 5 \times 17$  & 910 & 3672 & 17322 \\
    & $3 \times 85$           & 670 & 5316 & \textbf{15366} \\
    & $15 \times 17$          & 830 & 4072 & 16522 \\
    & $5 \times 51$           & 842 & 3655 & 16285 \\
    \hline
    511  & $7 \times 73$  & 1446 & 12238 & \textbf{36820} \\
    \hline
    \multirow{3}{*}{1023}
    &  $3 \times 11 \times 31$ & 4760 & 21198 & 111638\\
    & $11 \times 93$           & 50057 & 22672 & 118755\\
    & $33 \times 31$           & 4450 & 24174 & \textbf{108724}\\
    \hline
    2047 & $23 \times 89$ & 15204 & 77770 & \textbf{397054}\\
    \hline
    \multirow{10}{*}{4095}
    & $5 \times 7 \times 9 \times 13$ & 22690 & 81303 & 603173\\
    & $5 \times 7 \times 117$ & 18070 & 98488  & 514098\\
    & $5 \times 9 \times 91$  & 19450 & 99573  & 546923\\
    & $5 \times 13 \times 63$ & 20480 & 96838  & 567878\\
    & $7 \times 9 \times 65$  & 18910 & 91509  & 526439\\
    & $7 \times 13 \times 45$ & 21780 & 83305  & 584245\\
    & $9 \times 13 \times 35$ & 23860 & 88908  & 637688\\
    & $35 \times 117$    & 19240 & 106093 & 548613\\
    & $45 \times 91$     & 18540 & 101575 & 527995\\
    & $65 \times 63$     & 16700 & 107044 & \textbf{491144}\\
    \hline
  \end{tabular}
  \caption{The complexities of our PFCFTs  of
    $(2^l-1)$-point DFTs over GF$(2^l)$ ($4\le l \le 12$) for possible decompositions. The 31- and
    127-point DFTs are omitted since our PFCFTs reduce to CFFTs in
    these two cases.}
  \label{tab:PFA}
\end{table}

\subsection{Complexity reduction}
\label{sec:result}
We reduce the additive complexities of our PFCFTs in three
steps. First, we reduce the complexities of short cyclic
convolutions. Second, we use these short cyclic convolutions to
construct CFFTs of moderate length. Third, we use CFFTs of moderate
length as sub-DFTs to construct our PFCFTs.


Our first step is to obtain short cyclic convolutions with low
complexity. Suppose an $L$-point cyclic convolution
$\mathbf{a}^{(L)}\otimes \mathbf{b}^{(L)}$ is calculated with the
bilinear form
$\Q^{(L)}(\mathbf{R}^{(L)}\mathbf{a}^{(L)}\cdot\mathbf{P}^{(L)}\mathbf{b}^{(L)})$,
we apply the CSE algorithm to reduce the additive complexities
required in the multiplication with $\mathbf{P}^{(L)}$ and $\Q^{(L)}$
(the multiplication $\mathbf{R}^{(L)}\mathbf{a}^{(L)}$ is
precomputed). The additive complexities
$\mathcal{C}_\mathrm{CSE}(\Q^{(L)})$,
$\mathcal{C}_\mathrm{CSE}(\mathbf{P}^{(L)})$, and the total additive
complexity
$\mathcal{C}_\mathrm{CSE}(\mathbf{Q}^{(L)})+\mathcal{C}_\mathrm{CSE}(\mathbf{P}^{(L)})$
as well as the multiplicative complexities are listed in
Tab.~\ref{tab:convCmplx}.
The short cyclic convolution algorithms for lengths 2--9 and 11 are
from \cite{BlahutFast,BlahutECC,Private,Ning2009}, and the 10-point
cyclic convolution is built from 2- and 5-point convolutions while the
12-point cyclic convolution is built from 3- and 4-point
convolutions. Convolutions with longer lengths are not needed in this
paper.

The second step is to reduce the additive complexity of CFFTs with
moderate length, which will be used to build long DFTs.  Because of
their moderate lengths, we can run the CSE algorithm many times and
choose the best results. For any $k$ so that $k|2^l-1$ ($4 \le l \le
12$) and $k<200$, the multiplicative and reduced additive complexity
of the $k$-point CFFTs are shown in Tab.~\ref{tab:shortFFTs}.

\begin{table*}[!htb]
  \addtolength{\tabcolsep}{-1pt}
  \centering
  \begin{tabular}{|c|c|c|c|c|c|c|c|c|c|c|c|c|}
    \hline
    \multirow{2}{*}{$N$} & \multicolumn{3}{|c|}{\cite{Wang1988}} & \multicolumn{3}{|c|}{PFA\cite{Truong2006}}
    & \multicolumn{3}{|c|}{DCFFT \cite{Chen2009c,Ning2009}} & \multicolumn{3}{|c|}{PFCFT} \\
    \cline{2-13}
    & mult. & add. & total & mult. &
    add. & total & mult. & add. & total & mult. & add. & total \\
    \hline
    15   & 41    & 97    & 384    & --   & --    & --     & 16   & 74    & 186    & 21    & 81    & 221 \\
    63   & 801   & 801   & 9612   & --   & --    & --     & 97   & 759   & 1826   & 131   & 552   & 1993 \\
    255  & 1665  & 5377  & 30352  & 1135 & 3887  & 20902  & 586  & 6736  & 15526  & 670   & 5316  & 15366  \\
    511  & 13313 & 13313 & 239634 & 6516 & 17506 & 128278 & 1014 & 23130 & 40368  & 1446  & 12238 & 36820  \\
    1023 & 32257 & 32257 & 645140 & 5915 & 30547 & 142932 & 2827 & 75360 & 129073 & 4450  & 24174 & 108724 \\
    2047 & 76801 & 76801 & 1689622& --   & --    & --     & 7812 & 529720& 693772 & 15204 & 77770 & 397054 \\
    4095 & 180225& 180225& 4325400& --   & --    & --     & --   & --    & --     & 16700 & 107044 & 491144 \\
    \hline
  \end{tabular}
  \caption{Comparison of the DFT complexity reduction results in the literature and our paper.}
  \label{tab:comp}
\end{table*}

Two possible schemes can be used to reduce the additive complexity of
CFFTs in (\ref{eq:cfft}), and they may lead to different additive
complexities.  Scheme 1 reduces $\mathcal{C}(\mathbf{AQ})$, while
scheme 2 reduces $\mathcal{C}(\mathbf{A})$ and
$\mathcal{C}(\mathbf{Q})$ separately.  From a theoretical point of
view, it is easy to show that
$\mathcal{C}_\mathrm{opt}(\mathbf{AQ})\le\mathcal{C}_\mathrm{opt}(\mathbf{A})+\mathcal{C}_\mathrm{opt}(\mathbf{Q})$,
since $(\mathbf{AQ})\mathbf{x}=\mathbf{A}(\mathbf{Qx})$. However, this
property may not hold for the CSE algorithm since it is not able to
identify all the linearly dependent patterns in the matrix. We may
benefit from reducing $\mathcal{C}(\A)$ and $\mathcal{C}(\Q)$ for the
following two reasons. First, $\mathbf{\Q}$ has a block diagonal
structure, which is similar as $\mathbf{P}$, therefore we can find a
better reduction result for $\mathcal{C}(\Q)$. Second, the size of
$\mathbf{A}$ is smaller than $\mathbf{AQ}$, and hence the CSE
algorithm requires less memory and time to reduce $\mathbf{A}$ than to
reduce $\mathbf{AQ}$.  The additive complexities based on schemes 1
and 2 are both listed, and the boldface additive complexity is the
smaller one for each $k$.

In the third step, we use the CFFTs of moderate lengths in
Tab.~\ref{tab:shortFFTs} as sub-DFTs to construct long DFTs. Hence, we
use the complexities listed in Tab.~\ref{tab:shortFFTs} to derive the
computational complexity of the DFTs with composite lengths $2^{l}-1$
over GF$(2^l)$ for $4 \le l \le 12$. All the possible decomposition of
$2^l-1$ with factors less than 200 and the corresponding
multiplicative and additive complexities are listed in
Tab.~\ref{tab:PFA}. Note that for each sub-DFT, the scheme with the
smaller additive complexity listed in Tab.~\ref{tab:shortFFTs} is used
in our PFCFTs to reduce the total additive complexity.  Since some
lengths of the DFTs have more than one decomposition, it is possible
that one decomposition scheme has a smaller additive complexity but a
larger multiplicative complexity than another one. Take 4095-point DFT
as an example. The decomposition $7 \times 9 \times 65$ requires 91509
additions and 18910 multiplications, while the $7 \times 13 \times 45$
decomposition requires 21780 additions and 83305 multiplications.
Therefore a metric of the total complexity is needed to compare the
total complexities of different decompositions. In this paper, we
follow \cite{Chen2009c} and assume the complexity of a multiplication
over GF$(2^l)$ is $2l-1$ times of that of an addition over the same
field, and the total complexity of an DFT is a weighted sum of the
additive and multiplicative complexities, i.e.,
$\mathrm{total}=(2l-1)\times \mathrm{mult} +\mathrm{add}$. This
assumption is based on both software and hardware implementation
considerations \cite{Chen2009c}. Using this metric, in
Tab.~\ref{tab:PFA} the smallest total complexity for each DFT is in
boldface.

\subsection{Complexity comparison}
For composite $N= 2^l-1$ ($4\le l \le 12$), the complexities of our
PFCFTs are compared to the best DFTs in the literature known to us in
Tab.~\ref{tab:comp}.
Although the FFTs in \cite{Wang1988} are proved asymptotically fast,
the complexities of our PFCFTs are only a fraction of those in
\cite{Wang1988}. 
Compared with the previous PFA result \cite{Truong2006},
our PFCFTs have much smaller multiplicative complexities due to CFFTs
used for the sub-DFTs. The additive complexities of our PFCFTs for
$N=511$ and $1023$ are much smaller due to CSE. Thus our PFCFTs have
smaller total complexities than those in \cite{Truong2006}. Compared
with the direct CFFT (DCFFT) results in \cite{Chen2009c} and
\cite{Ning2009}, for $N \geq 63$, our PFCFTs have much smaller
additive complexities due to their decomposition structure. For
instance, the additive complexity of our PFCFTs is about half of that
of the DCFFT for $N=511$, and one third for $N=1023$. Although the
multiplicative complexities of our PFCFTs are somewhat larger than
DCFFTs, the reduced additive complexity outweighs the increased
multiplicative complexity for long DFTs.
Hence, our PFCFTs have smaller total complexities than CFFTs in \cite{Chen2009c} and \cite{Ning2009} for $N\ge 255$, and the
improvement increases as $N$ grows.

When the lengths of DFTs are prime (for example, 31-point DFT over
GF$(2^5)$, 127-point DFT over GF$(2^7)$, and $8191$-point DFT over
GF$(2^{13})$), our PFCFTs reduce to CFFTs. Therefore, our PFCFTs and
CFFTs have the same computational complexities in such cases.

\begin{figure}[!htb]
  \centering
  \includegraphics[width=8cm]{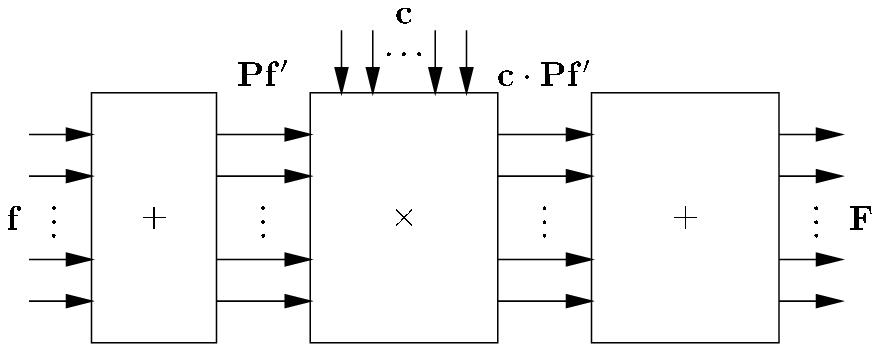}
  \caption{The circuitry of CFFTs.}
  \label{fig:CFFT}
\end{figure}

\section{Hardware Architecture of our PFCFTs}
\label{sec:arch}
CFFTs have a bilinear form, and therefore their hardware
implementation consists of three parts as shown in
Fig.~\ref{fig:CFFT}. The input vector $\mathbf{f}$ is first fed to an
pre-addition network, which reorders $\mathbf{f}$ into $\mathbf{f}'$
and then computes $\mathbf{Pf'}$. Then a multiplicative network
computes the entry-wise product of $\mathbf{c}$ and
$\mathbf{Pf}'$. The DFT $\mathbf{F}$ is finally computed by the
post-addition network which corresponds to the linear transform
$\mathbf{AQ}$. Although the structure in Fig.~\ref{fig:CFFT} appears
simple, the two additive networks are very complex for long DFTs. Even
with CSE, the two additive networks still require a large number of
additions. Furthermore, both lack regularity and structure, making it
difficult to implement them efficiently in hardware.

\begin{figure}[!htb]
  \centering
  \includegraphics[width=6.5cm]{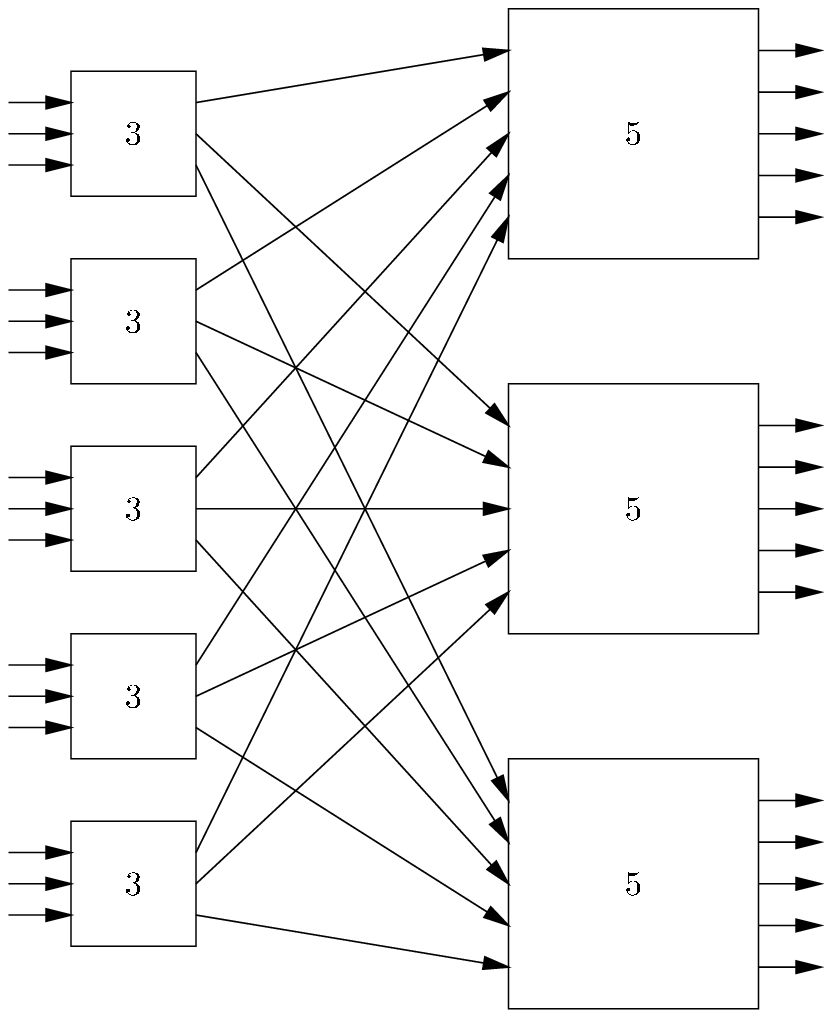}
  \caption{The regular structure of our 15-point PFCFT.}
  \label{fig:PFCFT3x5}
\end{figure}

In contrast, our PFCFTs are more suitable for hardware implementation
due to their regular structure.  Since long DFTs are decomposed into
short sub-DFTs, their hardware implementation becomes much easier and
can be reused in our PFCFTs. Fig.~\ref{fig:PFCFT3x5} illustrates the
regular structure of our 15-point PFCFT. Instead using the circuitry
in Fig.~\ref{fig:CFFT} for 15-point CFFTs, we only need to design a
3-point CFFT module and a 5-point CFFT module, and our 15-point PFCFT
is obtained by using these two modules, as shown in
Fig.~\ref{fig:PFCFT3x5}. Even when the total complexity of our PFCFTs
is higher than that of CFFTs, our PFCFTs may be considered due to
their advantage in hardware implementation.

\section{Conclusion}
\label{sec:conclusion}
In this paper, we propose a family of fast DFTs over GF$(2^l)$ ($4\le l \le 12$) with
composite lengths, called PFCFTs. Our PFCFTs have smaller total
complexities that previously proposed FFTs when $N\geq 255$. Our PFCFTs of very long lengths (say $4095$-point) are the only known efficient DFTs of such lengths.  Finally, our PFCFTs also have advantages in
hardware implementation due to their regular structure.

\bibliographystyle{IEEEtran}
\bibliography{IEEEabrv,FFT}

\end{document}